\documentclass{article}
\usepackage[utf8]{inputenc}
\usepackage{enumitem}
\usepackage{spconf,amsmath,graphicx,hyperref}
\usepackage{booktabs}
\usepackage{multirow}
\usepackage{url}
\usepackage{cite}
\usepackage{wrapfig}

\title{Speaking Clearly: A Simplified Whisper-based Codec for Low-Bitrate Speech Coding}
 
\name{Xin Zhang$^1$, Lin Li$^1$, Xiangni Lu$^1$, Jianquan Liu$^2$, Kong Aik Lee$^3$\thanks{Corresponding author: Lin Li (cathylilin@whut.edu.cn)}}

\address{$^1$School of Computer Science and Artificial Intelligence,Wuhan University of Technology, Wuhan, China \\
         $^2$NEC Corporation, Tokyo, Japan \\
         $^3$Department of Electrical and Electronic Engineering, The Hong Kong Polytechnic University, Hong Kong}

\begin{document}
\ninept

\maketitle

\begin{abstract}
Speech codecs serve as bridges between continuous speech signals and large language models, yet face an inherent conflict between acoustic fidelity and semantic preservation. To mitigate this conflict, prevailing methods augment acoustic codecs with complex semantic supervision. We explore the opposite direction: a semantic-first approach that starts from a semantically-capable model and adapts it for high-fidelity acoustic reconstruction. Through empirical analysis, we discover that targeted architectural simplification can unlock the acoustic modeling potential of Whisper, a text-aligned Automatic Speech Recognition (ASR) model. Based on this finding, we propose {\bf SimWhisper-Codec}, a novel codec that balances the semantic and acoustic preservation by leveraging a frozen, simplified Whisper encoder without requiring external supervision. Experimental results demonstrate that SimWhisper-Codec achieves superior performance in both semantic preservation and acoustic quality compared to semantically-supervised codecs such as Mimi Codec and SpeechTokenizer at similar bitrates, validating the effectiveness of our semantic-first approach. Code is available at \url{https://github.com/ZhangXinWhut/SimWhisper-Codec}.
\end{abstract}


\section{Introduction}
\label{sec:intro}

In recent years, Speech Large Language Models (Speech LLMs) have garnered significant attention from the research community, demonstrating remarkable performance across a range of tasks~\cite{adi2023generative,hassid2023textually,zhang2023speechgpt,fang2025llamaomni}. The success of Speech LLMs is critically underpinned by a core component: the speech codec. This component serves as a crucial bridge, converting continuous audio signals into discrete tokens suitable for LLM modeling\cite{borsos2023audiolm}, thereby connecting raw audio with the model's semantic understanding.

However, current speech codecs face an inherent conflict between the preservation of semantic content and acoustic fidelity, as optimizing for one typically degrades the other \cite{defossez2024moshi, defossezCSA23}. This trade-off is particularly pronounced at low bitrates, where achieving high performance in both dimensions remains difficult. To mitigate this conflict, prevailing methods augment acoustic-centric codecs with external semantic supervision through various strategies. For instance, SpeechTokenizer \cite{zhang2024speechtokenizer} guides the first residual vector quantization (RVQ) layer through semantic distillation from HuBERT \cite{hsu2021hubert}, while Mimi Codec in Moshi \cite{defossez2024moshi} employs split RVQ \cite{zeghidour2021soundstream} with SSL model \cite{chen2022wavlm} supervision on one quantization branch. PAST \cite{har2025past} incorporates auxiliary phonetic tasks such as phoneme classification and ASR, and XY-Tokenizer \cite{gong2025xy} employs multi-task learning with LLM-based ASR supervision and multi-stage training. While effective, these methods typically rely on complex semantic supervision.

\begin{figure}[!t]
  \begin{minipage}[b]{1.0\linewidth}
    \centering
    \centerline{\includegraphics[width=\linewidth]{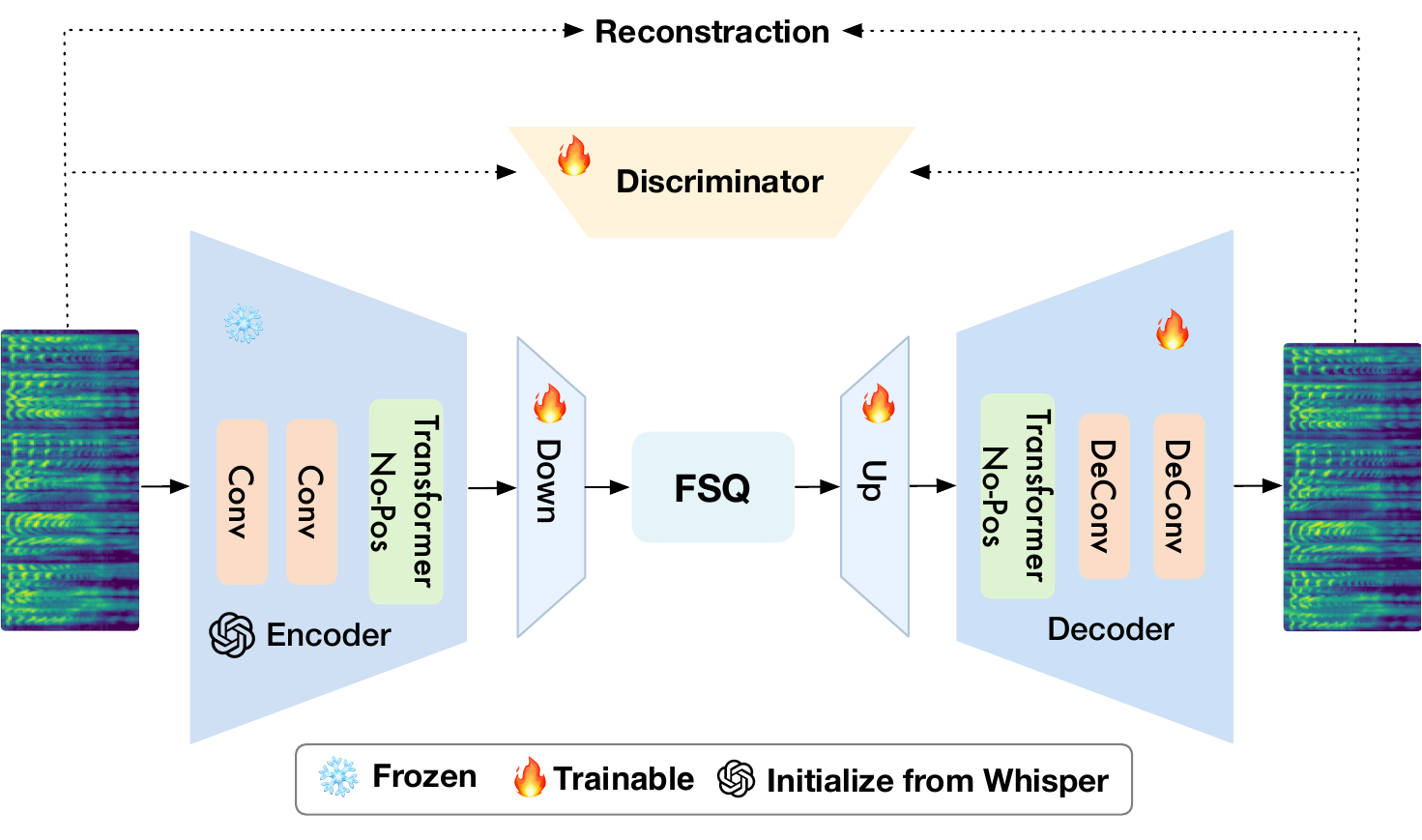}}

  \end{minipage}
  \caption{\textbf{Overview of SimWhisper-Codec.} A simplified Whisper encoder
  with FSQ discretization and a symmetric decoder. Model architecture
  and training procedure are detailed in Section~\ref{ssec:method}.}
  \label{fig:codec_overview}
  \vspace{-2mm}
\end{figure}

In this work, we explore the opposite direction: instead of enhancing acoustic codecs with semantic supervision, we start from Whisper \cite{radford2023robust}, a text-aligned ASR model, and adapt it for high-fidelity acoustic reconstruction. However, this adaptation encounters a task mismatch—ASR systems are designed to achieve invariance to acoustic variations for content extraction \cite{serdyuk2016invariant}, while acoustic reconstruction requires preserving fine-grained acoustic details for fidelity. To investigate this task mismatch, we conduct empirical analysis examining how different architectural components of Whisper affect its acoustic reconstruction capabilities. Through empirical analysis presented in Section~\ref{sec:analysis}, we discover that targeted architectural simplification—specifically removing the convolutional front-end nonlinearity (GELU activation) and absolute positional encodings \cite{vaswani2017attention}—substantially enhances the model's ability to preserve fine-grained acoustic information. Based on this finding, we propose SimWhisper-Codec (see Figure~\ref{fig:codec_overview}), a low-bitrate (1.1 kbps at 16 kHz) codec that combines a \emph{simplified} Whisper encoder, Finite Scalar Quantization (FSQ) \cite{mentzer2024finite}, and a symmetric decoder, enabling \emph{single-stage} training without semantic supervision.

Our contributions are as follows:
\begin{itemize}[itemsep=0.5em,topsep=0.5pt,parsep=0pt,partopsep=0pt,leftmargin=*]
  \item We propose SimWhisper-Codec, a novel codec that simultaneously models semantic and acoustic information through targeted architectural simplifications of Whisper's encoder combined with FSQ quantization and symmetric decoding, eliminating the need for external semantic supervision.
  \item Experimental results demonstrate that SimWhisper-Codec achieves state-of-the-art acoustic quality (3.29 PESQ-NB, 2.72 PESQ-WB) and competitive semantic preservation (2.75 WER) at similarly low bitrates without external supervision, validating the effectiveness of our approach.
\end{itemize}

\section{EMPIRICAL ANALYSIS OF COMPONENTS HINDERING ACOUSTIC RECONSTRUCTION}
\label{sec:analysis}

We conduct an empirical analysis to identify which architectural components in \emph{multilingual} Whisper encoders adversely affect acoustic reconstruction capabilities.

\subsection{Component Analysis and Hypotheses}
\label{ssec:theory}

\noindent\textbf{Convolutional Front-End Nonlinearity.} The Whisper encoder's front-end consists of two convolutional layers with GELU activation functions. While these nonlinear activations enable complex feature transformations beneficial for ASR tasks \cite{radford2023robust, gulati2020conformer}, we hypothesize that they suppress spectral details essential for acoustic reconstruction. By removing these activations, the convolutional layers become purely linear transformations that better preserve input signal structure and retain acoustic details necessary for reconstruction.

\noindent\textbf{Absolute Positional Encodings.} Absolute positional encodings assign fixed ``identity markers'' to each temporal position in the sequence \cite{kazemnejad2023impact, zhang2024empirical}. We hypothesize that absolute positional encodings are detrimental to acoustic reconstruction because: (1) acoustic features should remain position-invariant—a phoneme /a/ should have identical representation regardless of temporal location; (2) speech contains repetitive structures that absolute encodings differentiate, hindering pattern recognition for reconstruction. These theoretical considerations motivate our experimental validation.

\subsection{Validation Analysis}
\label{ssec:validation}

To validate our hypotheses, we conduct controlled analysis experiments using LJSpeech\footnote{\url{https://keithito.com/LJ-Speech-Dataset/}}. We extract frame-level hidden states from the final layer of each Whisper encoder variant, then condition identical HiFiGAN vocoders \cite{kong2020hifi} on these features to assess reconstruction quality. This setup allows us to isolate the impact of specific architectural components on acoustic modeling capability while keeping the vocoder constant. Crucially, all encoder variants remain \textbf{frozen} during HiFiGAN training to serve as feature extractors \cite{ChenZWLLD23}.

We examine the effect of removing each component individually, then their combined removal. The results confirm our hypotheses: removing convolutional front-end nonlinearity yields substantial improvements with PESQ-NB increasing from 1.24 to 3.60 (+2.36), STOI from 0.82 to 0.96 (+0.14), and SIM from 0.81 to 0.86 (+0.05). Removing absolute positional encodings also confirms this with PESQ-NB increasing from 1.24 to 2.95 (+1.71), STOI from 0.82 to 0.94 (+0.12), and SIM from 0.81 to 0.84 (+0.03). 

Simultaneously removing both components yields the best reconstruction performance: PESQ-NB reaches 3.67 (+2.43), STOI achieves 0.97 (+0.15), and SIM attains 0.87 (+0.06). This demonstrates complementary hindering effects—the nonlinearity suppresses spectral details while positional encodings interfere with flexible attention patterns. Table~\ref{tab:whisper_ablation} summarizes these results, with metrics defined in Section~\ref{ssec:metrics}. We validate that these improvements generalize across the entire Whisper model family, from tiny to large-v3, demonstrating the robustness of our findings.

\begin{figure}[t]
  \begin{minipage}[b]{.48\linewidth}
    \centering
    \centerline{\includegraphics[width=\linewidth]{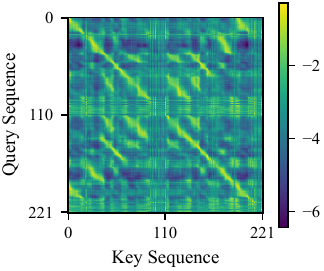}}
    \centerline{(a) With positional encoding}\medskip
  \end{minipage}
  \hfill
  \begin{minipage}[b]{.48\linewidth}
    \centering
    \centerline{\includegraphics[width=\linewidth]{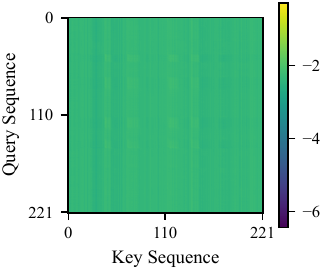}}
    \centerline{(b) Without positional encoding}\medskip
  \end{minipage}

  \vspace{-2mm}
  \caption{Self-attention maps (log-weights, head-averaged) from a middle Transformer layer for the utterance ``one, two, three, four, four, three, two, one''. The maps demonstrate how absolute positional encodings affect attention patterns in repetitive structures.}
  \label{fig:attn_maps}
  \vspace{-2mm}
\end{figure}

To further understand how positional encodings affect attention mechanisms, we visualize self-attention patterns from a middle Transformer layer. Figure~\ref{fig:attn_maps} provides compelling evidence for our positional encoding hypothesis: removing absolute positional encodings reduces diagonal dominance in self-attention patterns (0.031→0.005), enabling attention to spread across the sequence and reveal content-driven interactions around repeated segments.

\begin{table}[t]
  \centering
  \scriptsize
  \setlength{\tabcolsep}{3.2pt}\renewcommand{\arraystretch}{1.05}
  \caption{Reconstruction quality of Whisper encoder variants with HiFiGAN on LJSpeech.}
  \label{tab:whisper_ablation}
  \vspace{2pt}
  \begin{tabular}{lcccc}
    \toprule
    Variant & SIM $\uparrow$ & STOI $\uparrow$ & PESQ-NB $\uparrow$ & PESQ-WB $\uparrow$ \\
    \midrule
    Whisper encoder (baseline) & 0.81 & 0.82 & 1.24 & 1.14 \\
    \ \;– remove \textit{absolute} PEs only & 0.84 & 0.94 & 2.95 & 2.49 \\
    \ \;– remove \textit{both} stem GELUs only & 0.86 & 0.96 & 3.60 & 3.28 \\
    \ \;– remove both components & \textbf{0.87} & \textbf{0.97} & \textbf{3.67} & \textbf{3.33} \\
    \bottomrule
  \end{tabular}
  \vspace{-2mm}
\end{table}

Having established that simultaneously removing both components achieves optimal reconstruction performance, these findings provide the foundation for our codec design. The substantial improvements in acoustic reconstruction quality (+2.43 PESQ-NB, +0.15 STOI) demonstrate that architectural simplification can effectively unlock Whisper's potential for high-fidelity acoustic modeling. Based on these insights, we next present SimWhisper-Codec, which leverages the simplified Whisper encoder as a frozen feature extractor in a complete speech codec framework.

\vspace{-2mm}
\section{Method}
\label{ssec:method}

\subsection{SimWhisper-Codec}
\label{ssec:simwhisper-codec}

\textbf{Motivation.} Rather than augmenting acoustic codecs with external semantic supervision, we explore the opposite direction: starting from Whisper's inherent semantic capabilities and adapting it for high-quality acoustic reconstruction. The key insight is that Whisper's extensive multilingual training and text alignment provide natural semantic grounding, eliminating the need for additional semantic models. However, certain architectural components designed for ASR invariance may hinder fine-grained acoustic preservation. Based on our empirical findings, we propose SimWhisper-Codec, which employs a frozen simplified Whisper encoder paired with FSQ quantization and a symmetric trainable decoder. By leveraging Whisper's inherent semantic capabilities while enhancing its acoustic modeling through architectural simplification, our approach enables single-stage training without external semantic supervision.

\subsection{Model Architecture}
\label{ssec:architecture}

As shown in Figure \ref{fig:codec_overview}, the SimWhisper-Codec architecture is an end-to-end model comprising a simplified Whisper encoder, a downsampling module, a quantizer, an upsampling module, and a symmetric decoder. The downsampling module and quantizer collectively form an information bottleneck, compressing the encoder's output by reducing both temporal resolution and feature dimensionality.

\noindent\textbf{Encoder.} The encoder adopts the Whisper architecture initialized with
pre-trained weights, with two key modifications to enhance acoustic preservation. 
First, we remove the GELU non-linearities from the initial two convolutional layers
while preserving both the layer structure and learned weights from pre-training
to maintain compatibility with the pre-trained Whisper model. Second,
we completely remove the absolute positional encodings from the
Transformer blocks. This simplified encoder remains frozen
during codec training to serve as a powerful feature extractor, leveraging
the rich representations learned during Whisper's original ASR pre-training.

\noindent\textbf{Downsampler.} The downsampler reduces the temporal resolution by stacking
consecutive frames and aggregating temporal information into the channel
dimension. Subsequently, a series of residual blocks with dilated convolutions
and Snake activation functions \cite{ziyin2020neural} progressively compress the feature dimensionality
while capturing multi-scale temporal context.

\noindent\textbf{Quantizer.} We employ a Finite Scalar Quantization (FSQ) module \cite{mentzer2024finite}, which mitigates codebook collapse and obviates the need for complex training machinery such as exponential moving averages and commitment losses required by traditional VQ \cite{oord2017neural} methods.

\noindent\textbf{Upsampler.} The upsampler first reconstructs features through residual blocks with dilated convolutions and Snake activations. It then expands the channel dimension and unstacks the features to restore the original temporal resolution and dimensionality for the decoder.

\noindent\textbf{Decoder.} The decoder adopts a symmetric architecture to the encoder, with symmetry achieved by replacing the encoder's convolutional layers with transposed convolutions while maintaining the same architectural depth and feature dimensions. This design enables effective reversal of the encoding process to reconstruct mel-spectrogram representations from the upsampled features. A Vocos model \cite{siuzdak2024vocos} subsequently converts the spectral features to the final audio waveform.

\subsection{Training Objective} \label{ssec:training}

The codec is trained using a single-stage GAN-based approach. The generator minimizes the following composite loss function:

\begin{equation}
\mathcal{L}_{G} = \lambda_{\text{recon}}\mathcal{L}_{\text{recon}} + \lambda_{\text{adv}}\mathcal{L}_{\text{adv}} + \lambda_{\text{feat}}\mathcal{L}_{\text{feat}}
\label{eq:total_loss}
\end{equation}

\noindent where $\lambda_{\text{recon}}$, $\lambda_{\text{adv}}$, and $\lambda_{\text{feat}}$ control the weights of the multi-scale reconstruction loss $\mathcal{L}_{\text{recon}}$, adversarial loss $\mathcal{L}_{\text{adv}}$, and feature matching loss $\mathcal{L}_{\text{feat}}$, respectively.

\noindent\textbf{Multi-scale Reconstruction Loss.} We compute an L1 loss between the mel-spectrograms of the original and reconstructed audio across seven STFT scales. For each scale $k \in \{5,\dots, 11\}$, we calculate:

\begin{equation}
\mathcal{L}_{\text{recon}} = \sum_{k} \| M_k(x) - M_k(\hat{x}) \|_1
\label{eq:recon_loss}
\end{equation}

\noindent where $M_k(\cdot)$ denotes the mel-spectrogram computed with FFT size $2^k$, $x$ is the original audio, and $\hat{x}$ is the reconstructed audio.

\noindent\textbf{Adversarial Loss.} We employ a Least Squares GAN (LSGAN) objective \cite{mao2017least} to enhance perceptual quality. The discriminator loss is defined as:

\begin{equation}
\mathcal{L}_{D} = \frac{1}{N} \sum_{i=1}^{N} \left[ (D_i(x) - 1)^2 + D_i(G(z))^2 \right]
\label{eq:disc_loss}
\end{equation}

\noindent where $D_i$ represents the $i$-th discriminator output from (MPD \cite{kong2020hifi}, MS-STFTD \cite{defossezCSA23}), $N$ is the number of discriminators, and $G(z)$ is the generated audio. The generator adversarial loss is defined as:

\begin{equation}
\mathcal{L}_{\text{adv}} = \frac{1}{N} \sum_{i=1}^{N} (D_i(G(z)) - 1)^2
\label{eq:gen_adv}
\end{equation}

\noindent\textbf{Feature Matching Loss.} We compute an L1 loss between the feature maps of the discriminators for the real and generated audio, which prevents the generator from overtraining on the current discriminator and improves generation quality. The feature matching loss is formulated as:

\begin{equation}
\mathcal{L}_{\text{feat}} = \frac{1}{N \cdot K} \sum_{i=1}^{N} \sum_{j=1}^{K} \frac{\| D_i^j(x) - D_i^j(G(z)) \|_1}{\| D_i^j(x) \|_1 + \epsilon}
\label{eq:feat_loss}
\end{equation}

\noindent where $D_i^j(\cdot)$ denotes the $j$-th layer feature map from the $i$-th discriminator, $K$ is the number of feature layers, and $\epsilon$ is a small constant for numerical stability.

\section{Experiments}
\label{sec:experiments}

\subsection{Settings}
\label{ssec:settings}

\noindent\textbf{Dataset and Training Details.} We use the full training set of LibriSpeech \cite{panayotov2015librispeech} with 960 hours of speech data for training. The test-clean set with 2620 utterances is used for testing. All speech data are in 16 kHz with randomly cropped 2-second audio segments. Training is conducted on a single NVIDIA H100 GPU with batch size of 64 and gradient accumulation set to 1, resulting in an effective batch size of 64. The total training is performed for 1000,000 steps using a single-stage approach. Both generator and discriminator employ AdamW optimization with $\beta_1 = 0.8$, $\beta_2 = 0.99$, and weight decay $0.01$. A cosine annealing learning rate schedule is used, declining from $1 \times 10^{-4}$ to $0$ with 10k warmup steps for both components.

\noindent\textbf{Framework Configuration.} Our framework employs a symmetric encoder-decoder architecture. Both components are 12-layer Transformers based on the Whisper-small architecture, with 768-dimensional hidden states and 12 attention heads. The model processes 16~kHz audio, extracting 50~Hz feature sequences using a 25~ms window and a 10~ms hop size. The \textbf{downsampler} reduces the temporal resolution to 12.5~Hz via 4$\times$ frame stacking and compresses the feature dimension from 768 to 32 using residual blocks with Snake activations \cite{ziyin2020neural} and multi-scale dilated convolutions (dilations: 1, 3, 9). The \textbf{upsampler} employs a symmetric architecture, first passing through residual blocks and then restoring the 50~Hz resolution and 768 dimensions via unstacking. Following \cite{casanova2025low}, we use a Finite Scalar Quantization (FSQ) module configured with eight codebooks, four dimensions per code, and levels of [8, 7, 6, 6] to achieve a 1.1~kbps bitrate. The decoder reconstructs mel-spectrograms from upsampled features, which are then synthesized to 16~kHz waveforms by a 24-layer Vocos model with hop size 160.

\noindent\textbf{Baselines.} We compare our codec against representative
baselines at similar bitrates: EnCodec \cite{defossezCSA23} (1.5~kbps), DAC-RVQ3 \cite{kumar2023high} (1.5~kbps),
SpeechTokenizer \cite{zhang2024speechtokenizer} (1.0~kbps), Mimi-RVQ8 \cite{defossez2024moshi} (1.1~kbps), BigCodec(1.04~kbps), XCodec2.0 \cite{ye2025llasa} (0.8~kbps) and XY-Tokenizer \cite{gong2025xy} (1.0~kbps), all using official checkpoints.

\subsection{Evaluation Metrics}
\label{ssec:metrics}

Two aspects are evaluated: (i) \emph{acoustic reconstruction quality} of the synthesized audio, and (ii) \emph{semantic alignment} between the codec and text. All metrics are reported on \textbf{LibriSpeech test-clean} \cite{panayotov2015librispeech}.

\noindent\textbf{PESQ-WB/NB.} Wideband and narrowband PESQ are reported
following ITU-T P.862.2 and P.862, respectively \cite{rix2001pesq}.
Signals are resampled to 16~kHz (WB) and 8~kHz (NB) before scoring.

\noindent\textbf{STOI.} Short-Time Objective Intelligibility is used to measure
intelligibility correlation between reference and reconstructed signals
\cite{taal2011stoi}. Audio is resampled to 16~kHz prior to computation.

\noindent\textbf{SIM (speaker similarity).} Cosine similarity is computed
between speaker embeddings extracted from reference and reconstructed
utterances using a \emph{WavLM}-based speaker verification model\footnote{\url{https://github.com/microsoft/UniSpeech/tree/main/downstreams/speaker_verification}}.

\noindent\textbf{WER via external ASR.} Reconstructed audio is transcribed with HuBERT
large\footnote{\url{https://huggingface.co/facebook/hubert-large-ls960-ft}} and word error rate (WER) is computed against LibriSpeech references.

\begin{table}[t]
  \centering
  \scriptsize
  \setlength{\tabcolsep}{1.1pt}
  \renewcommand{\arraystretch}{1.05}
  \caption{Low-bitrate codec comparison. Bold indicates the best performance for each metric.}
  \label{tab:main_results_wide}
  \vspace{1mm}
  \begin{tabular}{l c c c c c c c c}
    \toprule
    & & Frame & \multicolumn{2}{c}{Semantic} & \multicolumn{4}{c}{Acoustic} \\
    \cmidrule(lr){4-5} \cmidrule(lr){6-9}
    Model & BPS & Rate & Semantic & WER & SIM & STOI & PESQ- & PESQ- \\
    & & & Supervision & $\downarrow$ & $\uparrow$ & $\uparrow$ & NB $\uparrow$ & WB $\uparrow$ \\
    \midrule
    Ground Truth & -- & -- & -- & 2.16 & 1.00 & 1.00 & 4.55 & 4.64 \\
    \midrule
    EnCodec \cite{defossezCSA23} & 1.5k & 75 & No & 5.62 & 0.60 & 0.85 & 1.94 & 1.56 \\
    DAC-RVQ3 \cite{kumar2023high} & 1.5k & 75 & No & 7.80 & 0.45 & 0.76 & 1.82 & 1.43 \\
    SpeechTokenizer \cite{zhang2024speechtokenizer} & 1.0k & 50 & Yes & 4.21 & 0.37 & 0.70 & 1.42 & 1.15 \\
    BigCodec \cite{xin2024bigcodec} & 1.04k & 80 & No & 2.92 & 0.84 & \textbf{0.93} & 3.26 & 2.68 \\
    Mimi-RVQ8 \cite{defossez2024moshi} & 1.1k & 12.5 & Yes & 3.24 & 0.73 & 0.90 & 2.79  & 2.24 \\
    XCodec2.0 \cite{ye2025llasa} & 0.8k & 50 & Yes & 2.61 & 0.82 & 0.92 & 3.04 & 2.43 \\
    XY-Tokenizer \cite{gong2025xy} & 1.0k & 12.5 & Yes & \textbf{2.46} & \textbf{0.85} & 0.92 & 3.10 & 2.50 \\
    SimWhisper-Codec (ours) & 1.1k & 12.5 & No & 2.75 & 0.83 & \textbf{0.93} & \textbf{3.29} & \textbf{2.72} \\
    \bottomrule
  \end{tabular}
  \vspace{-4mm}
\end{table}

\vspace{-2mm}
\subsection{Experimental Results}
\label{ssec:results}

Table~\ref{tab:main_results_wide} presents a comprehensive comparison of SimWhisper-Codec against representative baselines. Our approach effectively models both semantic and acoustic information simultaneously, validating our architectural simplification.

\noindent\textbf{Semantic Preservation.} SimWhisper-Codec achieves a competitive WER of \textbf{2.75}, comparable to strong supervised baselines like XY-Tokenizer (2.46) and XCodec2.0 (2.61). Notably, this is achieved \emph{without semantic supervision}, unlike these baselines which rely on complex distillation or multi-task learning. This suggests that our architectural simplification effectively preserves Whisper's semantic alignment capabilities.

\noindent\textbf{Acoustic Quality.} For acoustic reconstruction, SimWhisper-Codec delivers state-of-the-art performance. It achieves \textbf{3.29} PESQ-NB and \textbf{2.72} PESQ-WB. These scores surpass the leading acoustic-only baseline, BigCodec (3.26 and 2.68). It also matches BigCodec's high intelligibility (\textbf{0.93} STOI). Furthermore, with a speaker similarity (SIM) score of \textbf{0.83}, it demonstrates good preservation of speaker identity. These results validate our hypothesis: targeted architectural simplification unlocks Whisper's acoustic modeling potential.

\vspace{0mm}
\subsection{Ablation Study}
\label{ssec:ablation}

To validate the impact of removing different architectural components on codec performance, we conduct ablation studies on SimWhisper-Codec. For training simplicity, we maintain symmetric encoder-decoder architectures throughout all variants.

\begin{table}[t]
  \centering
  \scriptsize
  \setlength{\tabcolsep}{0.9pt}
  \renewcommand{\arraystretch}{1.05}
  \caption{Ablation study on SimWhisper-Codec architectural components. For training simplicity, encoder and decoder maintain symmetric architectures.}
  \label{tab:ablation_simwhisper}
  \vspace{2mm}
  \begin{tabular}{lccccc}
    \toprule
    & \multicolumn{1}{c}{Semantic} & \multicolumn{4}{c}{Acoustic} \\
    \cmidrule(lr){2-2} \cmidrule(lr){3-6}
    Variant & WER $\downarrow$ & SIM $\uparrow$ & STOI $\uparrow$ & PESQ-NB $\uparrow$ & PESQ-WB $\uparrow$ \\
    \midrule
    Whisper encoder (baseline) & 5.95 & 0.78 & 0.85 & 1.95 & 1.68 \\
    \ \;– remove \textit{absolute} PEs only & 5.42 & 0.80 & 0.87 & 2.34 & 1.96 \\
    \ \;– remove \textit{both} stem GELUs only & 3.74 & 0.81 & 0.89 & 2.51 & 2.10 \\
    \textbf{Ours: remove both components} & \textbf{2.75} & \textbf{0.83} & \textbf{0.93} & \textbf{3.29} & \textbf{2.72} \\
    \bottomrule
  \end{tabular}
  \vspace{-4mm}
\end{table}

Table~\ref{tab:ablation_simwhisper} demonstrates that both architectural modifications contribute to performance improvements. Removing GELU activations from the convolutional front-end has a more substantial impact on semantic preservation (WER: 5.95→3.74), while removing positional encodings provides moderate gains across all metrics. The combination of both modifications yields the best results, achieving a WER of 2.75, SIM of 0.83, and PESQ-WB of 2.72, validating our architectural design choices.

\vspace{-2mm}
\subsection{Preservation of Acoustic Attributes}
\label{ssec:acoustic_preservation}

The ablation study demonstrates that our architectural modifications yield superior codec performance. To further validate that the simplified encoder retains fine-grained acoustic cues essential for high-quality synthesis, we conduct a layer-wise probing experiment analyzing its preservation of pitch information. 

\begin{wrapfigure}[12]{r}{0.45\linewidth}
  \vspace{-8pt}
  \centering
  \includegraphics[width=0.95\linewidth]{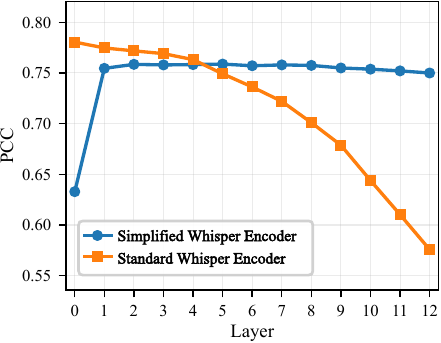}
  \vspace{-3pt}
  \caption{Pitch tracking performance.}
  \label{fig:f0_preservation}
  \vspace{-10pt}
\end{wrapfigure}

We train ridge regression models to predict frame-level fundamental frequency ($F_0$) from hidden states extracted from each encoder layer using THCHS-30 \cite{wang2015thchs}, a tonal Mandarin dataset with rich prosodic variation. Ground truth $F_0$ values are extracted using Parselmouth \cite{jadoul2018introducing}, evaluating only voiced frames with Pearson correlation coefficient (PCC). 

As shown in Figure~\ref{fig:f0_preservation}, the results reveal that simplified Whisper maintains stable $F_0$ tracking (PCC $\approx 0.76$) across all layers, while standard Whisper degrades from layer 6 onward (0.78→0.58). This indicates that our architectural modifications better preserve prosodic information essential for high-quality speech synthesis, providing additional evidence that the simplified encoder successfully retains acoustic details while maintaining semantic capabilities.

\vspace{-3mm}
\section{Conclusion}
\label{sec:conclusion}

We presented SimWhisper-Codec, a low-bitrate speech codec that mitigates the semantic-acoustic conflict through architectural simplification rather than complex supervision. We removed convolutional front-end nonlinearity and absolute positional encodings from the frozen Whisper encoder. This architectural simplification achieves superior acoustic quality while maintaining competitive semantic preservation. Results demonstrate that such simplification of Whisper can be more effective than semantic supervision approaches for speech codec design.

\bibliographystyle{IEEEbib}
\bibliography{refs}

\begin{thebibliography}{10}

\bibitem{adi2023generative}
R.~Algayres, Y.~Adi, T.~A. Nguyen, et~al.,
\newblock ``Generative spoken language model based on continuous word-sized audio tokens,''
\newblock in {\em EMNLP}, 2023, pp. 3008--3028.

\bibitem{hassid2023textually}
M.~Hassid, T.~Remez, T.~A. Nguyen, et~al.,
\newblock ``Textually pretrained speech language models,''
\newblock in {\em NeurIPS}, 2023.

\bibitem{zhang2023speechgpt}
D.~Zhang, S.~Li, X.~Zhang, et~al.,
\newblock ``Speechgpt: Empowering large language models with intrinsic cross-modal conversational abilities,''
\newblock in {\em Findings of EMNLP}, 2023, pp. 15757--15773.

\bibitem{fang2025llamaomni}
Q.~Fang, S.~Guo, Y.~Zhou, Z.~Ma, S.~Zhang, and Y.~Feng,
\newblock ``Llama-omni: Seamless speech interaction with large language models,''
\newblock in {\em ICLR}, 2025.

\bibitem{borsos2023audiolm}
Z.~Borsos, R.~Marinier, D.~Vincent, et~al.,
\newblock ``Audiolm: {A} language modeling approach to audio generation,''
\newblock {\em {IEEE} {ACM} Trans. Audio Speech Lang. Process.}, vol. 31, pp. 2523--2533, 2023.

\bibitem{defossez2024moshi}
A.~D{\'{e}}fossez, L.~Mazar{\'{e}}, M.~Orsini, et~al.,
\newblock ``Moshi: a speech-text foundation model for real-time dialogue,''
\newblock {\em arXiv preprint arXiv:2410.00037}, 2024.

\bibitem{defossezCSA23}
A.~D{\'{e}}fossez, J.~Copet, G.~Synnaeve, and Y.~Adi,
\newblock ``High fidelity neural audio compression,''
\newblock {\em Trans. Mach. Learn. Res.}, vol. 2023, 2023.

\bibitem{zhang2024speechtokenizer}
X.~Zhang, D.~Zhang, S.~Li, Y.~Zhou, and X.~Qiu,
\newblock ``Speechtokenizer: Unified speech tokenizer for speech language models,''
\newblock in {\em ICLR}, 2024.

\bibitem{hsu2021hubert}
W.-N. Hsu, B.~Bolte, Y.-H.~H. Tsai, K.~Lakhotia, R.~Salakhutdinov, and A.~Mohamed,
\newblock ``Hubert: Self-supervised speech representation learning by masked prediction of hidden units,''
\newblock {\em {IEEE} {ACM} Trans. Audio Speech Lang. Process.}, vol. 29, pp. 3451--3460, 2021.

\bibitem{zeghidour2021soundstream}
N.~Zeghidour, A.~Luebs, A.~Omran, J.~Skoglund, and M.~Tagliasacchi,
\newblock ``Soundstream: An end-to-end neural audio codec,''
\newblock {\em {IEEE} {ACM} Trans. Audio Speech Lang. Process.}, vol. 30, pp. 495--507, 2022.

\bibitem{chen2022wavlm}
S.~Chen, C.~Wang, Z.~Chen, et~al.,
\newblock ``Wavlm: Large-scale self-supervised pre-training for full stack speech processing,''
\newblock {\em {IEEE} J. Sel. Top. Signal Process.}, vol. 16, pp. 1505--1518, 2022.

\bibitem{har2025past}
N.~Har{-}Tuv, O.~Tal, and Y.~Adi,
\newblock ``{PAST:} phonetic-acoustic speech tokenizer,''
\newblock in {\em Interspeech}, 2025.

\bibitem{gong2025xy}
Y.~Gong, L.~Jin, R.~Deng, et~al.,
\newblock ``Xy-tokenizer: Mitigating the semantic-acoustic conflict in low-bitrate speech codecs,''
\newblock {\em arXiv preprint arXiv:2506.23325}, 2025.

\bibitem{radford2023robust}
A.~Radford, J.~W. Kim, T.~Xu, G.~Brockman, C.~McLeavey, and I.~Sutskever,
\newblock ``Robust speech recognition via large-scale weak supervision,''
\newblock in {\em ICML}, 2023, pp. 28492--28518.

\bibitem{serdyuk2016invariant}
D.~Serdyuk, K.~Audhkhasi, P.~Brakel, B.~Ramabhadran, S.~Thomas, and Y.~Bengio,
\newblock ``Invariant representations for noisy speech recognition,''
\newblock {\em arXiv preprint arXiv:1612.01928}, 2016.

\bibitem{vaswani2017attention}
A.~Vaswani, N.~Shazeer, N.~Parmar, et~al.,
\newblock ``Attention is all you need,''
\newblock in {\em NeurIPS}, 2017, pp. 5998--6008.

\bibitem{mentzer2024finite}
F.~Mentzer, D.~Minnen, E.~Agustsson, and M.~Tschannen,
\newblock ``Finite scalar quantization: {VQ}-{VAE} made simple,''
\newblock in {\em ICLR}, 2024.

\bibitem{gulati2020conformer}
A.~Gulati, J.~Qin, C.-C. Chiu, et~al.,
\newblock ``Conformer: Convolution-augmented transformer for speech recognition,''
\newblock in {\em Interspeech}, 2020, pp. 5036--5040.

\bibitem{kazemnejad2023impact}
A.~Kazemnejad, I.~Padhi, K.~N. Ramamurthy, P.~Das, and S.~Reddy,
\newblock ``The impact of positional encoding on length generalization in transformers,''
\newblock in {\em NeurIPS}, 2023.

\bibitem{zhang2024empirical}
Q.~Zhang, M.~Ge, H.~Zhu, et~al.,
\newblock ``An empirical study on the impact of positional encoding in transformer-based monaural speech enhancement,''
\newblock in {\em ICASSP}, 2024, pp. 1001--1005.

\bibitem{kong2020hifi}
J.~Kong, J.~Kim, and J.~Bae,
\newblock ``Hifi-gan: Generative adversarial networks for efficient and high fidelity speech synthesis,''
\newblock in {\em NeurIPS}, 2020.

\bibitem{ChenZWLLD23}
H.~Chen, H.~Zhang, L.~Wang, K.~A. Lee, M.~Liu, and J.~Dang,
\newblock ``Self-supervised audio-visual speaker representation with co-meta learning,''
\newblock in {\em ICASSP}, 2023, pp. 1--5.

\bibitem{ziyin2020neural}
L.~Ziyin, T.~Hartwig, and M.~Ueda,
\newblock ``Neural networks fail to learn periodic functions and how to fix it,''
\newblock in {\em NeurIPS}, 2020.

\bibitem{oord2017neural}
A.~van~den Oord, O.~Vinyals, and K.~Kavukcuoglu,
\newblock ``Neural discrete representation learning,''
\newblock in {\em NeurIPS}, 2017, pp. 6306--6315.

\bibitem{siuzdak2024vocos}
H.~Siuzdak,
\newblock ``Vocos: Closing the gap between time-domain and fourier-based neural vocoders for high-quality audio synthesis,''
\newblock in {\em ICLR}, 2024.

\bibitem{mao2017least}
X.~Mao, Q.~Li, H.~Xie, R.~Y.~K. Lau, Z.~Wang, and S.~P. Smolley,
\newblock ``Least squares generative adversarial networks,''
\newblock in {\em ICCV}, 2017, pp. 2813--2821.

\bibitem{panayotov2015librispeech}
V.~Panayotov, G.~Chen, D.~Povey, and S.~Khudanpur,
\newblock ``Librispeech: An {ASR} corpus based on public domain audio books,''
\newblock in {\em ICASSP}, 2015, pp. 5206--5210.

\bibitem{casanova2025low}
E.~Casanova, R.~Langman, P.~Neekhara, et~al.,
\newblock ``Low frame-rate speech codec: a codec designed for fast high-quality speech {LLM} training and inference,''
\newblock in {\em ICASSP}, 2025, pp. 1--5.

\bibitem{kumar2023high}
R.~Kumar, P.~Seetharaman, A.~Luebs, I.~Kumar, and K.~Kumar,
\newblock ``High-fidelity audio compression with improved {RVQGAN},''
\newblock in {\em NeurIPS}, 2023.

\bibitem{ye2025llasa}
Z.~Ye, X.~Zhu, C.-M. Chan, et~al.,
\newblock ``Llasa: Scaling train-time and inference-time compute for llama-based speech synthesis,''
\newblock {\em arXiv preprint arXiv:2502.04128}, 2025.

\bibitem{rix2001pesq}
A.~W. Rix, J.~G. Beerends, M.~P. Hollier, and A.~P. Hekstra,
\newblock ``Perceptual evaluation of speech quality (pesq)-a new method for speech quality assessment of telephone networks and codecs,''
\newblock in {\em ICASSP}, 2001, pp. 749--752.

\bibitem{taal2011stoi}
C.~H. Taal, R.~C. Hendriks, R.~Heusdens, and J.~Jensen,
\newblock ``An algorithm for intelligibility prediction of time-frequency weighted noisy speech,''
\newblock {\em {IEEE} Trans. Speech Audio Process.}, vol. 19, pp. 2125--2136, 2011.

\bibitem{xin2024bigcodec}
D.~Xin, X.~Tan, S.~Takamichi, and H.~Saruwatari,
\newblock ``Bigcodec: Pushing the limits of low-bitrate neural speech codec,''
\newblock {\em arXiv preprint arXiv:2409.05377}, 2024.

\bibitem{wang2015thchs}
D.~Wang and X.~Zhang,
\newblock ``{THCHS-30} : {A} free chinese speech corpus,''
\newblock {\em arXiv preprint arXiv:1512.01882}, 2015.

\bibitem{jadoul2018introducing}
Y.~Jadoul, B.~Thompson, and B.~de~Boer,
\newblock ``Introducing parselmouth: {A} python interface to praat,''
\newblock {\em J. Phonetics}, vol. 71, pp. 1--15, 2018.

\end{thebibliography}

\end{document}